\documentclass{cimento}

\usepackage{graphicx}
\usepackage{siunitx}
\usepackage{caption}
\usepackage{subcaption}
\usepackage{upgreek}
\usepackage{url}

\usepackage{xcolor}

\DeclareSIUnit{\rad}{rad}

\title{Characterization of Stitched Prototypes Chip for the ALICE ITS3 Upgrade}

\author{Michele Rignanese, on behalf of the ALICE collaboration}
\instlist{
\inst{} University of Padova, Dipartimento di Fisica e Astronomia "G.Galilei", Via Marzolo 8, 35131, Padova (PD), Italy 
\inst{} Istituto Nazionale di Fisica Nucleare, Sezione di Padova, Via Marzolo 8, 35131, Padova (PD), Italy
}

\begin{document}
\maketitle

\begin{abstract}
During LHC Long Shutdown 3, the ALICE experiment will replace the three innermost layers of its Inner Tracking System (ITS2) with a new vertex detector, the ITS3. This new detector will be assembled using wafer-scale, stitched Monolithic Active Pixel Sensors (MAPS) fabricated using a 65nm CMOS technology node, which will be thinned and bent to form truly cylindrical layers around the beam pipe. To validate the new technology, several prototypes were developed and extensively characterized. This work focuses on the results of a test beam campaign performed at the CERN PS in September 2024, using a \SI{10}{\giga\eV} pion beam, to estimate detection efficiency and spatial resolution of the babyMOSS prototype, a smaller version of the MOnolithic Stitched Sensor (MOSS). Both non-irradiated and irradiated chips are tested, and the results confirm that the prototypes meet the ITS3 requirements, demonstrating a detection efficiency above 99\%, with a fake-hit rate below 10$^{-6}$ hits/pixel/event and a spatial resolution around \SI{5}{\micro\meter}.
\end{abstract}

\section{Introduction}
\label{sec:introduction}
ALICE (A Large Ion Collider Experiment) is one of the four experiments of the CERN Large Hadron Collider (LHC), specialized in the study of Quark-Gluon Plasma produced in heavy-ion collisions. The current detector used for the reconstruction of tracks, collision and particle-decay vertices is the Inner Tracking System 2 (ITS2), which consists of seven concentric layers made up of thousands of ALPIDE chips, for a total silicon area of \SI{10}{\meter\squared} \cite{alpide, its2_upgrade}.
During Long Shutdown 3, the three innermost layers of ITS2 (\textit{i.e.} the Inner Barrel) will be upgraded with ITS3 to enhance the detector's performance, specifically the reconstruction of low p$_{\mathrm{T}}$ particles \cite{its3}.

\begin{figure}[ht]
    \centering
    \includegraphics[width=0.4\textwidth]{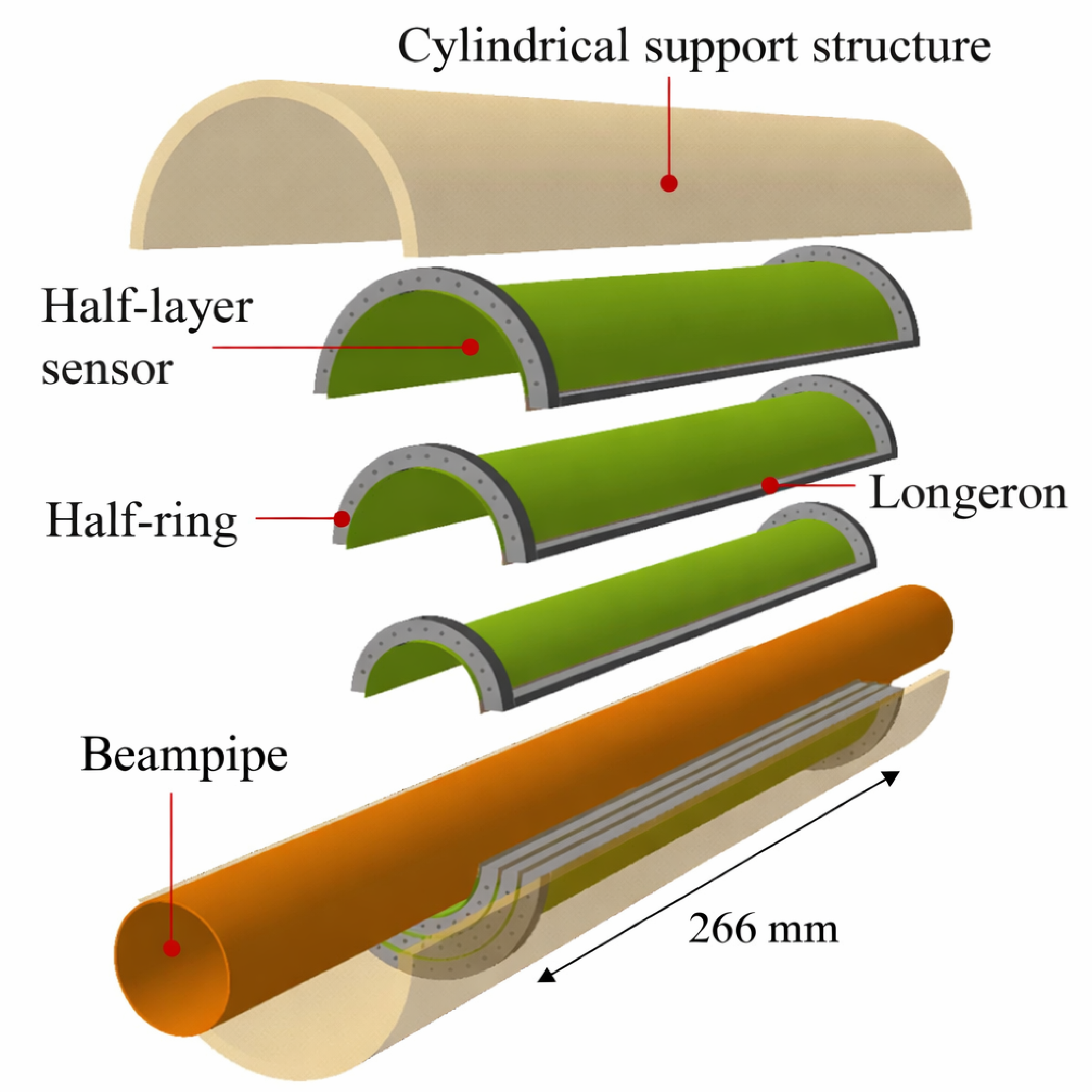}
    \caption{Schematic representation of the ITS3. Each layer is constituted by two wafer-scale sensors, bent around the beam pipe. The distance of the three layers from the beam pipe is of 19, 25.2 and 31.5 \si{\milli\meter}, respectively.}
    \label{fig:its3}
\end{figure}

The ITS3 concept relies on the fabrication of large-area stitched sensors bent to cylindrical shape, fabricated with a 65 nm CMOS imaging process. Thanks to stitching, it is possible to produce wafer-scale devices, which can be thinned down to approximately \SI{50}{\micro\meter} and bent around the beam pipe. As shown in figure \ref{fig:its3}, each half layer of the ITS3 will be held in place by carbon foam, eliminating the need for heavy support structures and drastically reducing the material budget from the current 0.36\% X$_0$ per layer to an average of 0.09\% X$_0$ per layer.

The successful implementation of ITS3 requires the validation of both the stitching process and the radiation of the new technology process used. The ITS3 radiation requirements are set to $4\times10^{12}$ \SI{1}{\mega\eV} n$_{eq}$/\si{\centi\meter\squared} Non-Ionizing Energy Loss (NIEL), and \SI{400}{\kilo\rad} Total Ionizing Dose (TID). 
The suitability of the 65 nm CMOS process was first demonstrated during the Multi-Layer Reticle 1 (MLR1) phase. During MLR1, small-scale prototypes were produced, such as the Analogue Pixel Test Structure (APTS) and the Digital Pixel Test Structure (DPTS), which were used to characterize the properties of the new technology used \cite{apts, dpts}. After the MLR1 phase, the R\&D campaign progressed to Engineering Run 1 (ER1) to demonstrate the feasibility of stitching. In ER1 both full-scale prototypes, the MOnolithic Stitched Sensor (MOSS) and the MOnolithic Stitched sensor with Timing (MOST), as well as smaller test structures known as babyMOSS and babyMOST were produced.
This work focuses on the characterization of the babyMOSS chip, presenting results on detection efficiency and spatial resolution obtained during a test beam at the CERN Proton Synchrotron (PS) carried out in September 2024.

\section{Stitched sensors: MOSS and babyMOSS}
\label{sec:stitched_sensors}
\begin{figure}[ht]
    \centering
    \includegraphics[width=0.8\textwidth]{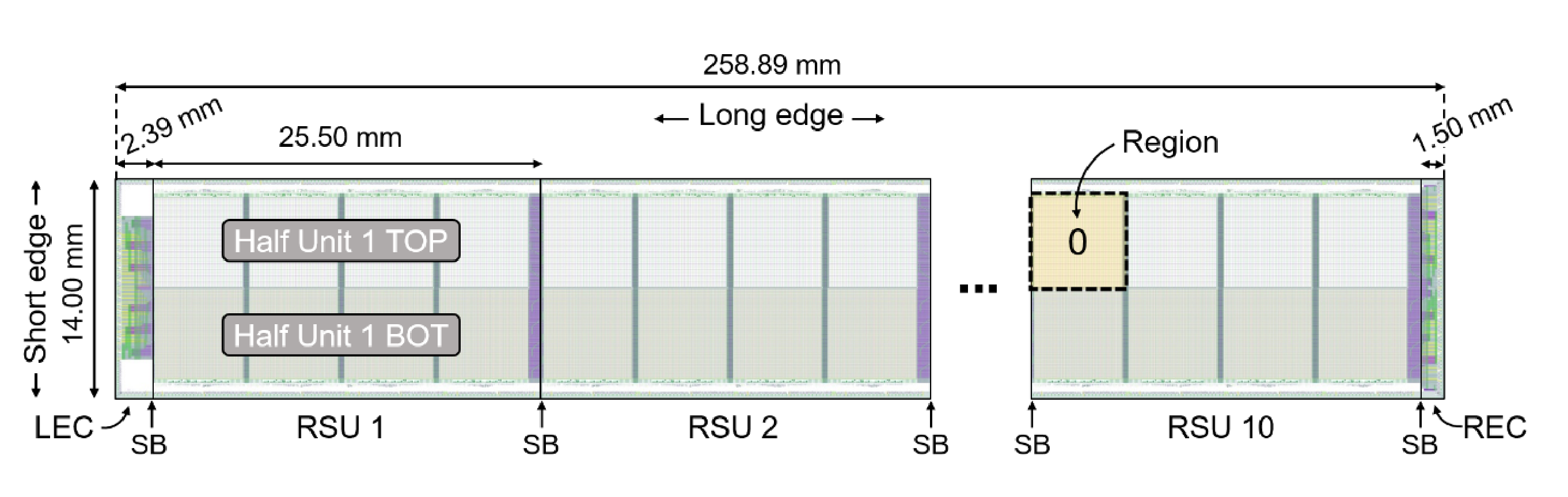}
    \caption{Schematic representation of the MOSS chip \cite{moss}. }
    \label{fig:moss}
\end{figure}

In order to realize wafer-scale sensors, the stitching technique is used to overcome the limitations of standard photolithography processes. In manufacturing of standard reticle-size sensors, the maximum chip size is limited by the reticle field of the photolithography mask, typically around $2.5\times3$ \si{\centi\meter\squared}. To produce large-area sensors required for the ITS3, stitching technique is used: multiple reticle size units are \textit{combined} during the wafer exposure. This process allows for the manufacturing of large-area sensors by repeating the fundamental building block, known as Repeated Sensor Unit (RSU).
The largest prototype used to define the layout and the design parameters for the final chip is the MOSS chip \cite{moss}. The MOSS chip has an active area of $26\times1.4$ \si{\centi\meter\squared}, and it is made by repeating ten RSUs together with the Left and the Right End Caps. A schematic representation of the MOSS chip is shown in figure \ref{fig:moss}.

\begin{figure}[ht]
    \centering
    \begin{subfigure}{0.48\textwidth}
        \centering
        \includegraphics[width=\textwidth]{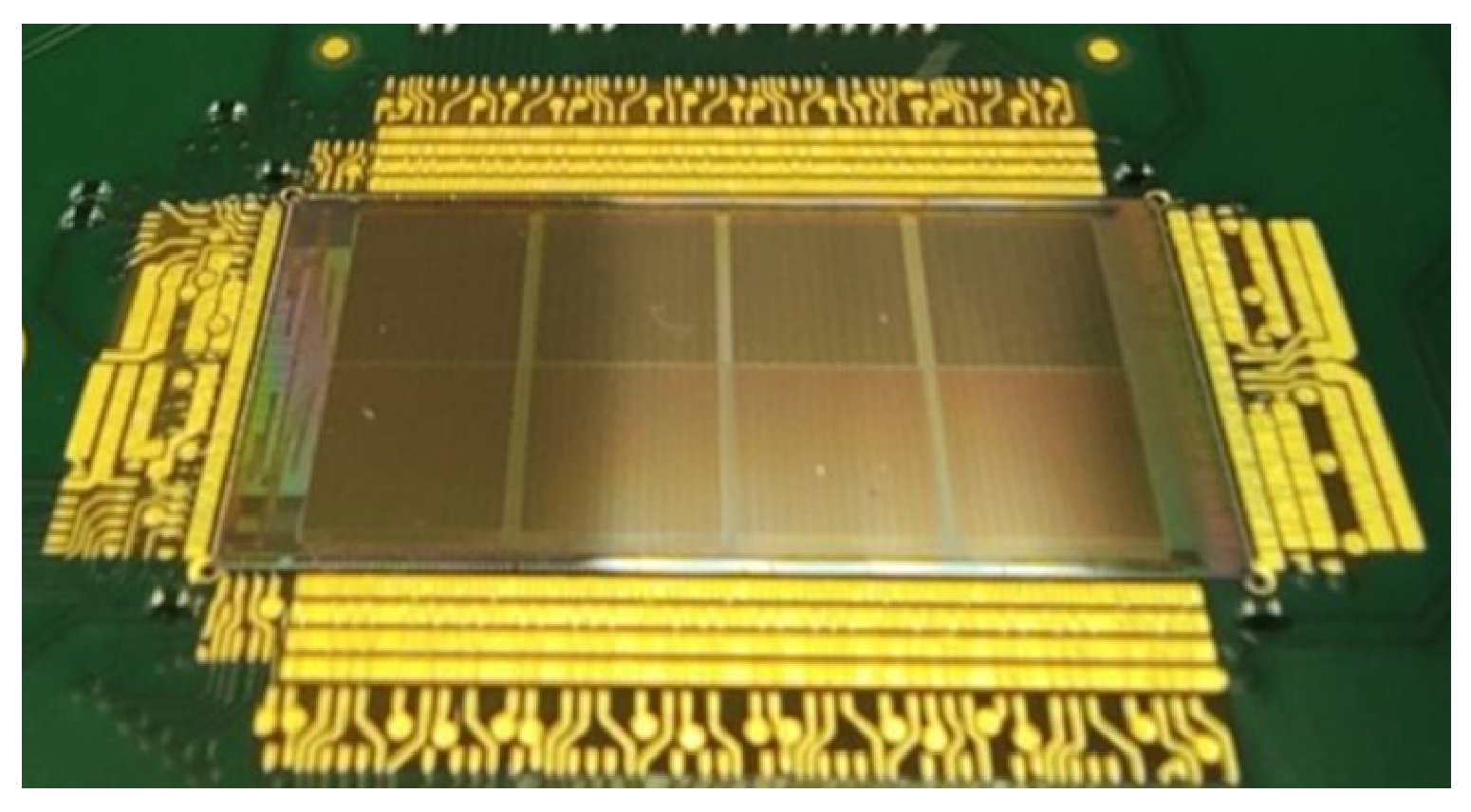}
        \caption{}
        \label{fig:babyMOSS}
    \end{subfigure}
    \hfill
    \begin{subfigure}{0.48\textwidth}
        \centering
        \includegraphics[width=\textwidth]{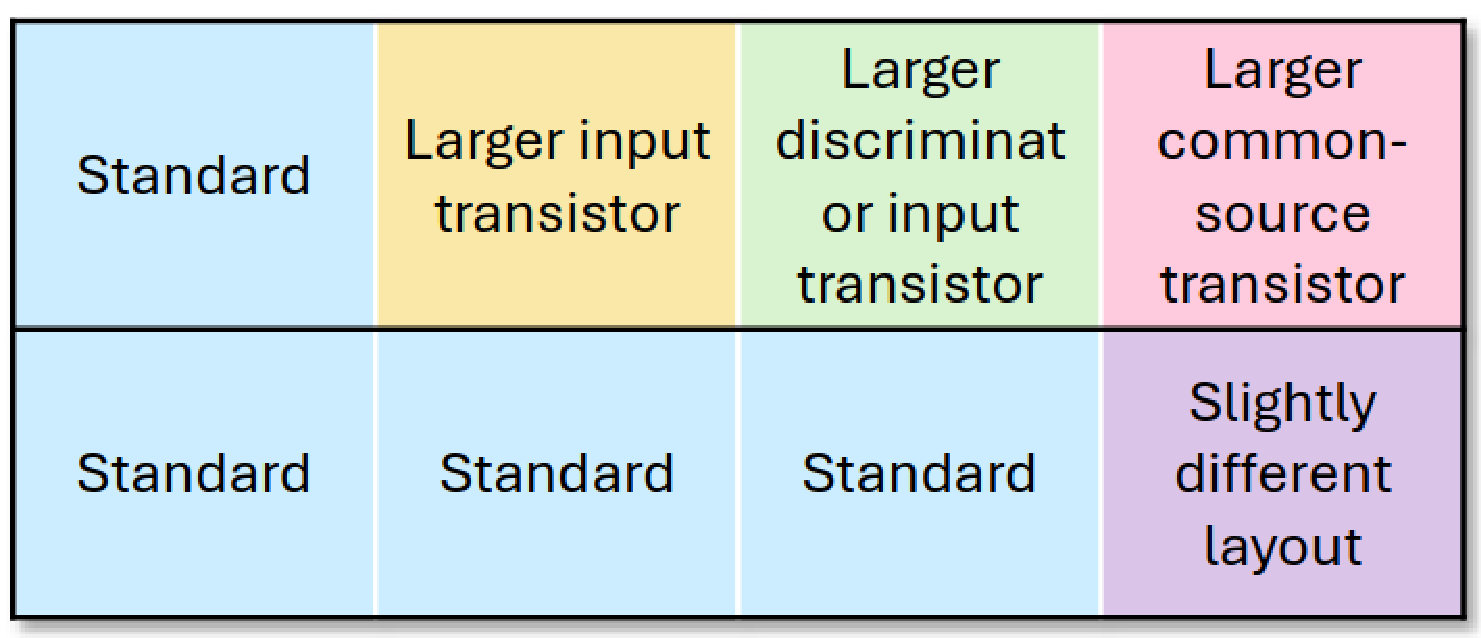}
        \caption{}
        \label{fig:babyMOSS_regions}
    \end{subfigure}
    \caption{BabyMOSS chip bonded on its carrier board (\subref{fig:babyMOSS}). Different front-end flavours implemented in all the eight regions (\subref{fig:babyMOSS_regions}).}
\end{figure}

The babyMOSS chip is one MOSS RSU and a Right End Cap. The pixel matrix of the babyMOSS is divided in two half-units (HUs) and each HU is further divided in four regions, implementing different front-end electronic flavours. The pixel matrix and the different front-end flavours are shown in figure \ref{fig:babyMOSS} and figure \ref{fig:babyMOSS_regions}, respectively. The Top HU features a pixel pitch of \SI{22.5}{\micro\meter}, while the Bottom HU implement smaller pixels of \SI{18}{\micro\meter}. This layout allows for a direct comparison of performance as a function of pixel pitch and front-end flavours within the same silicon die.

\section{Test beam setup}
\label{sec:setup}
\begin{figure}[ht]
    \begin{subfigure}{0.48\textwidth}
        \centering
        \includegraphics[width=0.65\textwidth]{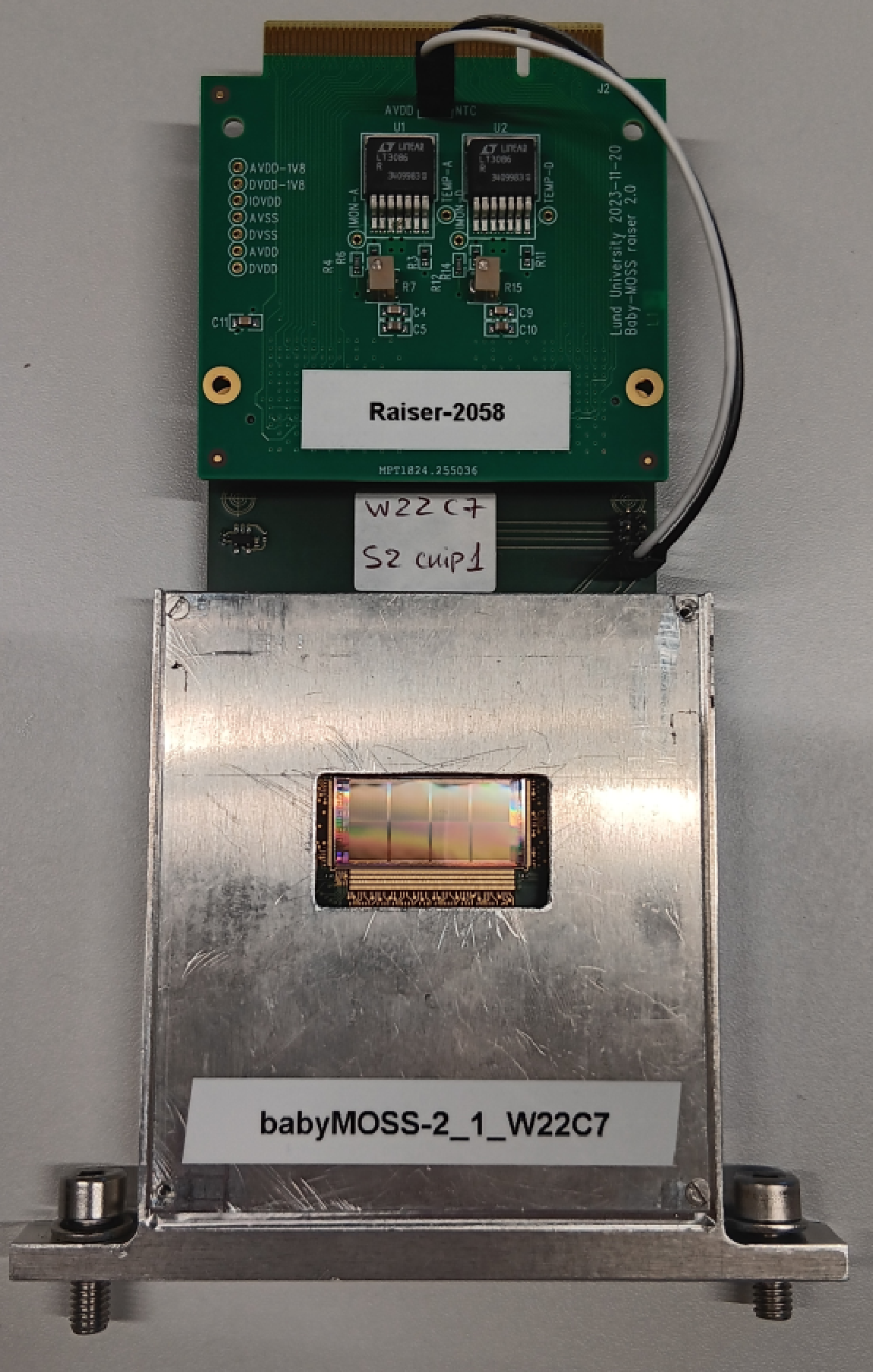}
        \caption{}
        \label{fig:babyMOSS_support}
    \end{subfigure}
    \hfill
    \begin{subfigure}{0.48\textwidth}
        \centering
        \includegraphics[width=0.75\textwidth]{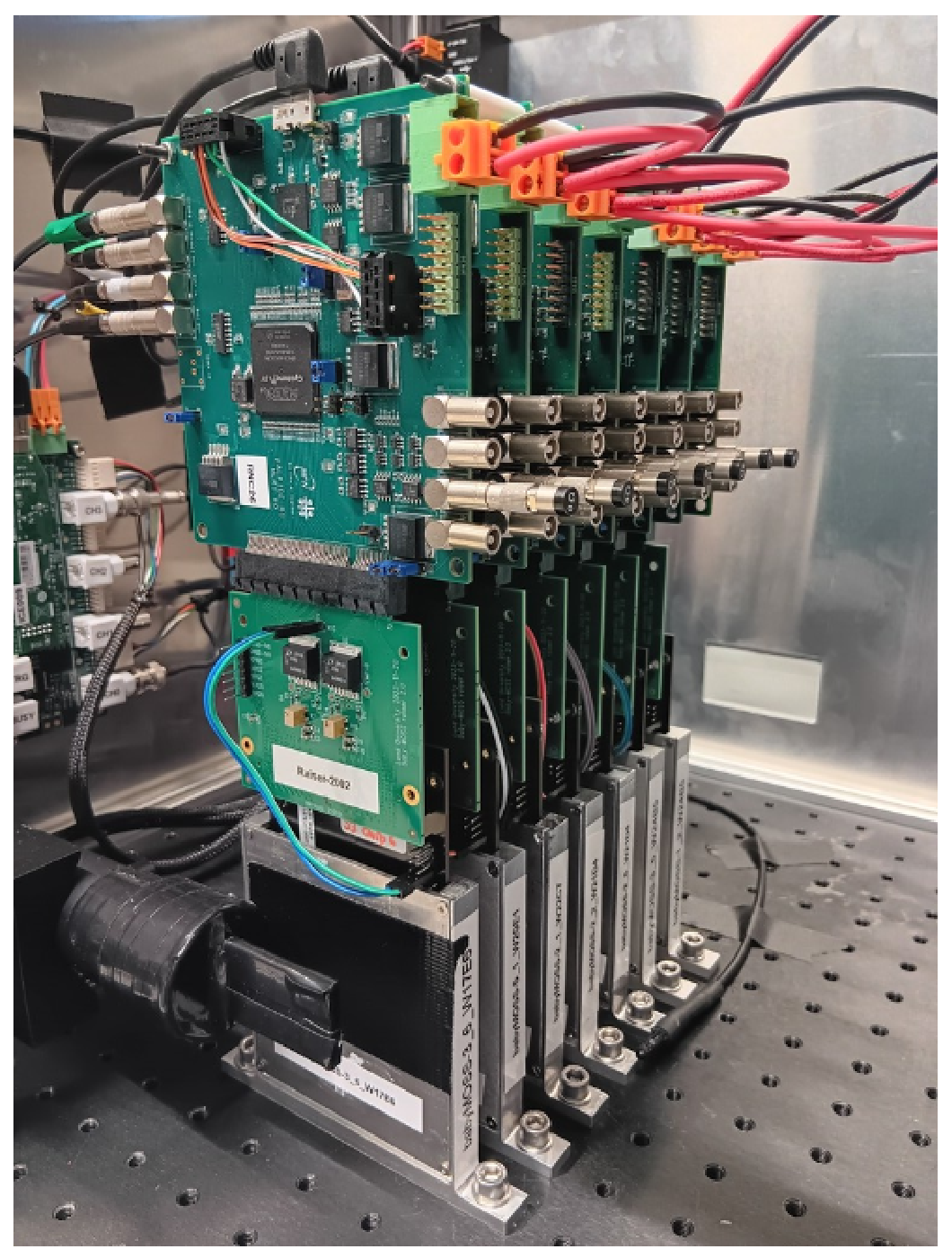}
        \caption{}
        \label{fig:telescope}
    \end{subfigure}
    \caption{BabyMOSS chip bonded on its carrier board inside test beam support (\subref{fig:babyMOSS_support}). BabyMOSS telescope with PMT (\subref{fig:telescope}). The distance between the planes is set to \SI{2.5}{\centi\meter}. Each babyMOSS is readout by the \textit{DAQ-Raiser} system, which is constituted by the DAQ board and the Raiser board acting as an interface between the DAQ board and the carrier board, where the chip is bonded.}
\end{figure}

To validate the performance of the babyMOSS chip in terms of detection efficiency and spatial resolution, several test beam campaigns were conducted. In what follows, only measurements made at the CERN PS facility in September 2024 using a \SI{10}{GeV} $\uppi^-$ beam are discussed. Measurements were made using a babyMOSS telescope (see figure \ref{fig:telescope}).
The experimental setup consists of six babyMOSS tracking planes, and of the Device Under Test (DUT) is placed in the middle. The trigger signal is generated by the coincidence of two scintillators coupled to photomultiplier tubes (PMTs). More details on the babyMOSS telescope and trigger logic can be found in \cite{panasenko_talk}.

Three DUTs were tested during the test beam: a non-irradiated DUT, the babyMOSS-2\_1\_W22C7, and two irradiated ones, babyMOSS-2\_2\_W02F4 and babyMOSS-3\_3\_W02F4. These two devices were irradiated to a NIEL of 10$^{13}$ 1 \si{\mega\eV} n$_{eq}$/\si{\centi\meter\squared}, exceeding the radiation dose expected during ITS3 operation. The MOSS sensor is designed with a low-dose n-type blanket implanted beneath the collection electrode, with gaps at the pixel edges. Wafers with gaps of different width (i.e. variations, called wafer splits) have been produced. For the Top HU the gaps have a width of \SI{2.5}{\micro\meter} and \SI{5}{\micro\meter} for split 1 and split 2 wafers, respectively. For the Bottom HU instead, the gaps are \SI{2.5}{\micro\meter} wide for both splits. The two irradiated samples are extracted from the same split 1 wafer, while the non-irradiated one comes from one split 2 wafer.

\section{Results}
\label{sec:results}
In this section analysis results of detection efficiency and spatial resolution of the babyMOSS-2\_1\_W22C7 and of babyMOSS-2\_2\_W02F4 are presented. 

Data reconstruction and analysis were performed using the Corryvreckan analysis framework \cite{corryvreckan}. The analysis workflow include noisy-pixels masking, clusterization of hits for each plane, telescope pre-alignment and alignment with track reconstruction, and finally alignment of all planes including the DUT.
After all alignment phases, the DUT performance is evaluated considering:

\begin{itemize}
    \item Detection efficiency and Fake-Hit Rate (FHR)
    \item Spatial resolution and average cluster size
\end{itemize}

\subsection{Efficiency and Fake-Hit Rate}
\label{subsec:eff_fhr}
\begin{figure}[ht]
    \centering
    \begin{subfigure}{0.49\textwidth}
        \centering
        \includegraphics[width=\textwidth]{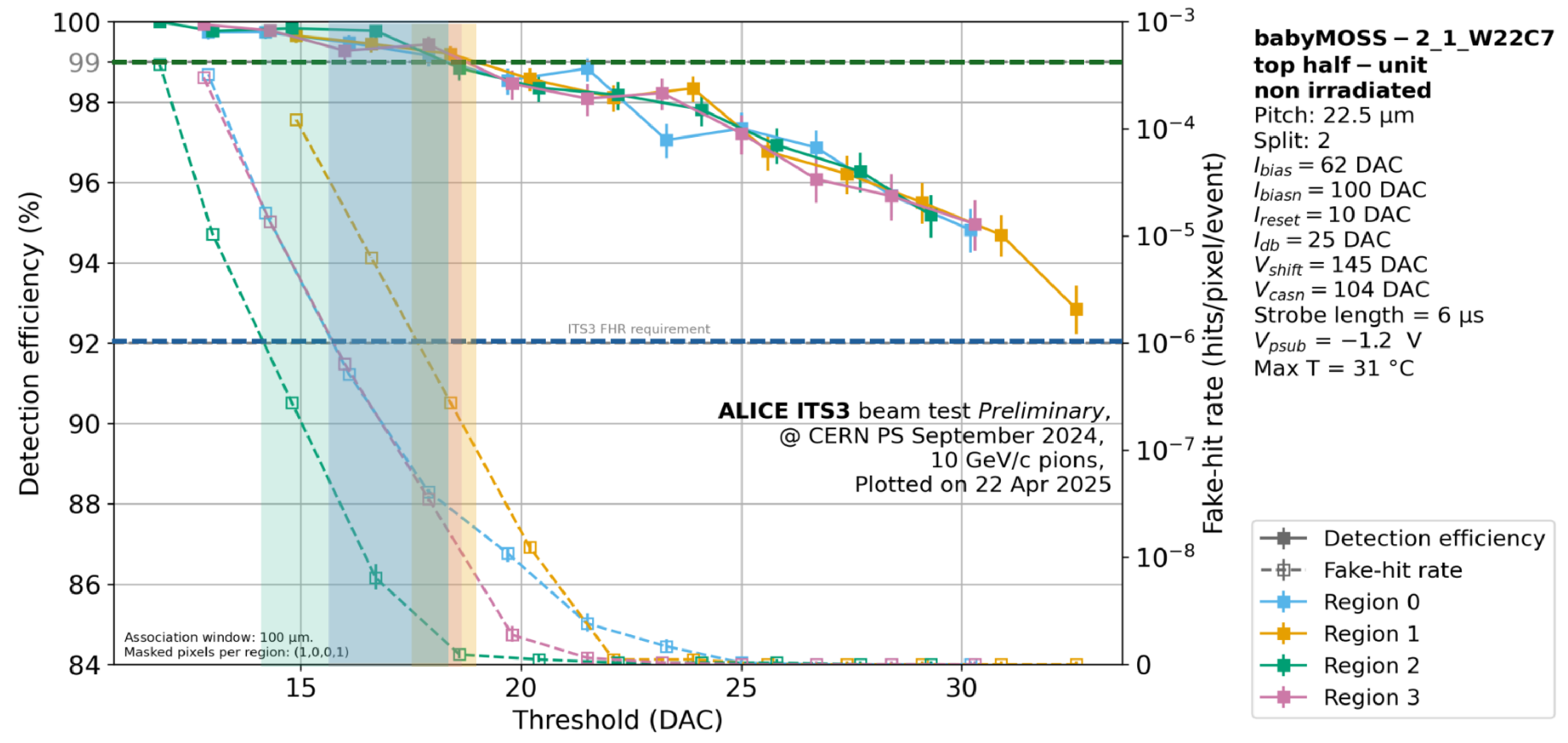}
        \caption{}
        \label{fig:eff_W22C7_top}
    \end{subfigure}
    \hfill
    \begin{subfigure}{0.49\textwidth}
        \centering
        \includegraphics[width=\textwidth]{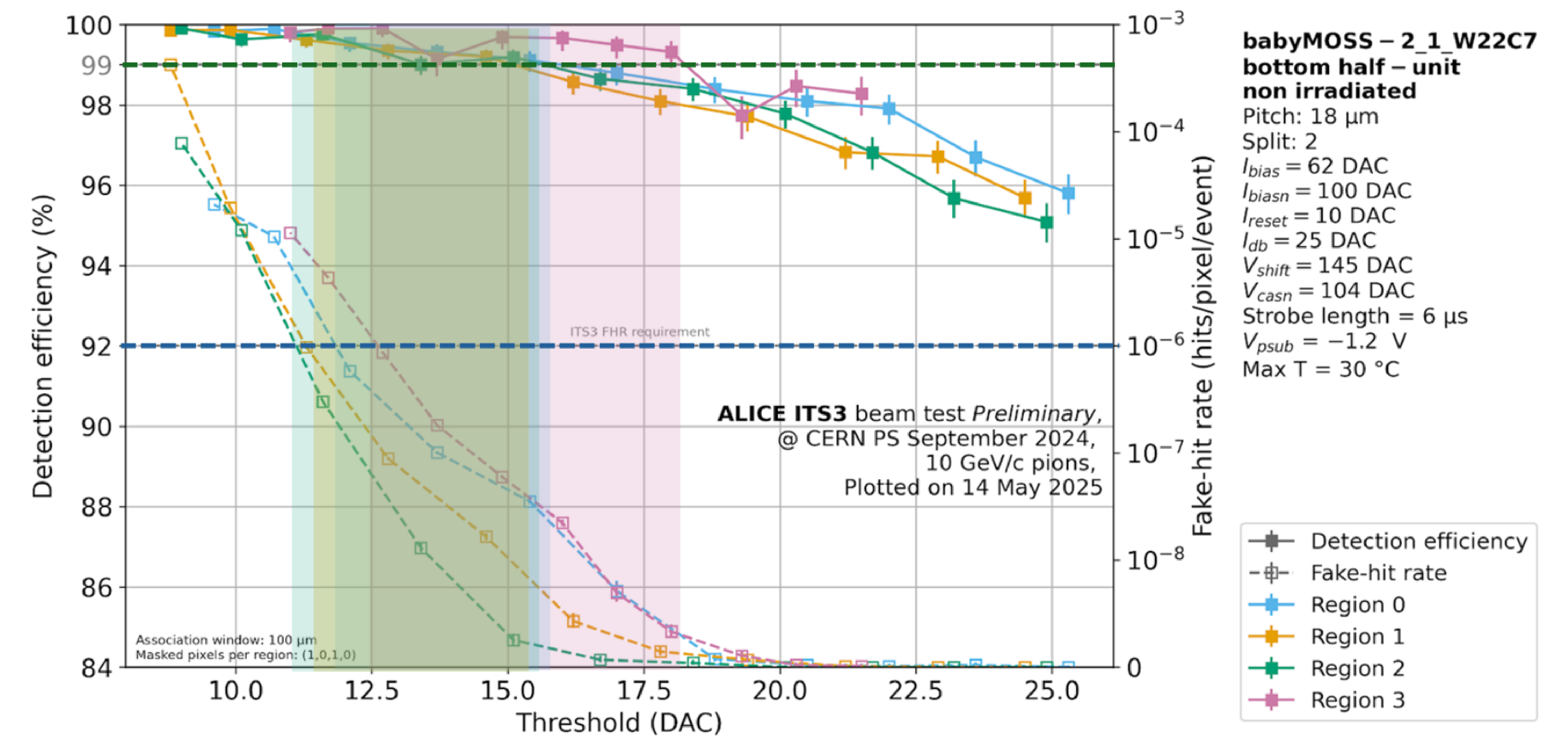}
        \caption{}
        \label{fig:eff_W22C7_bottom}
    \end{subfigure}
    \caption{Detection efficiency and FHR for the non-irradiated device for Top HU (\subref{fig:eff_W22C7_top}) and Bottom HU (\subref{fig:eff_W22C7_bottom}). The threshold ranges in which efficiency is above 99\% and FHR below $10^{-6}$ hits/pixel/event are highlighted for each region.}
    \label{fig:eff_W22C7}
\end{figure}

The detection efficiency and the FHR are evaluated as a function of the threshold, expressed in DAC units. Efficiency is computed as the ratio of tracks with an associated hit on the DUT (using an association window of \SI{100}{\micro\meter}) over the total number of tracks. The FHR, instead, is measured in the same conditions with the beam off. Results for the non-irradiated babyMOSS are shown in figure \ref{fig:eff_W22C7_top} and in figure \ref{fig:eff_W22C7_bottom} for the Top and Bottom HU, respectively.

\begin{figure}[ht]
    \centering
    \begin{subfigure}{0.49\textwidth}
        \centering
        \includegraphics[width=\textwidth]{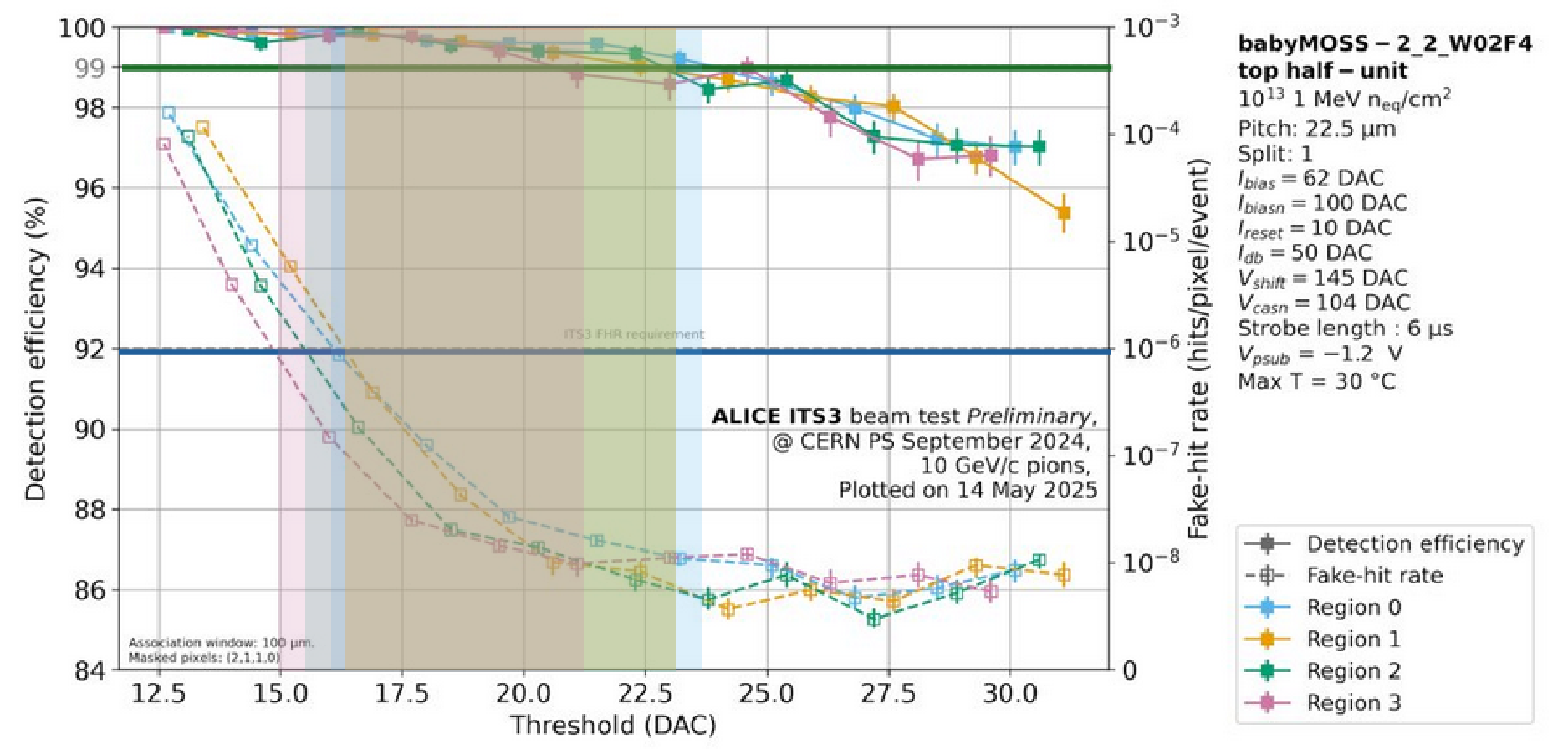}
        \caption{}
        \label{fig:eff_W02F4_top}
    \end{subfigure}
    \hfill
    \begin{subfigure}{0.49\textwidth}
        \centering
        \includegraphics[width=\textwidth]{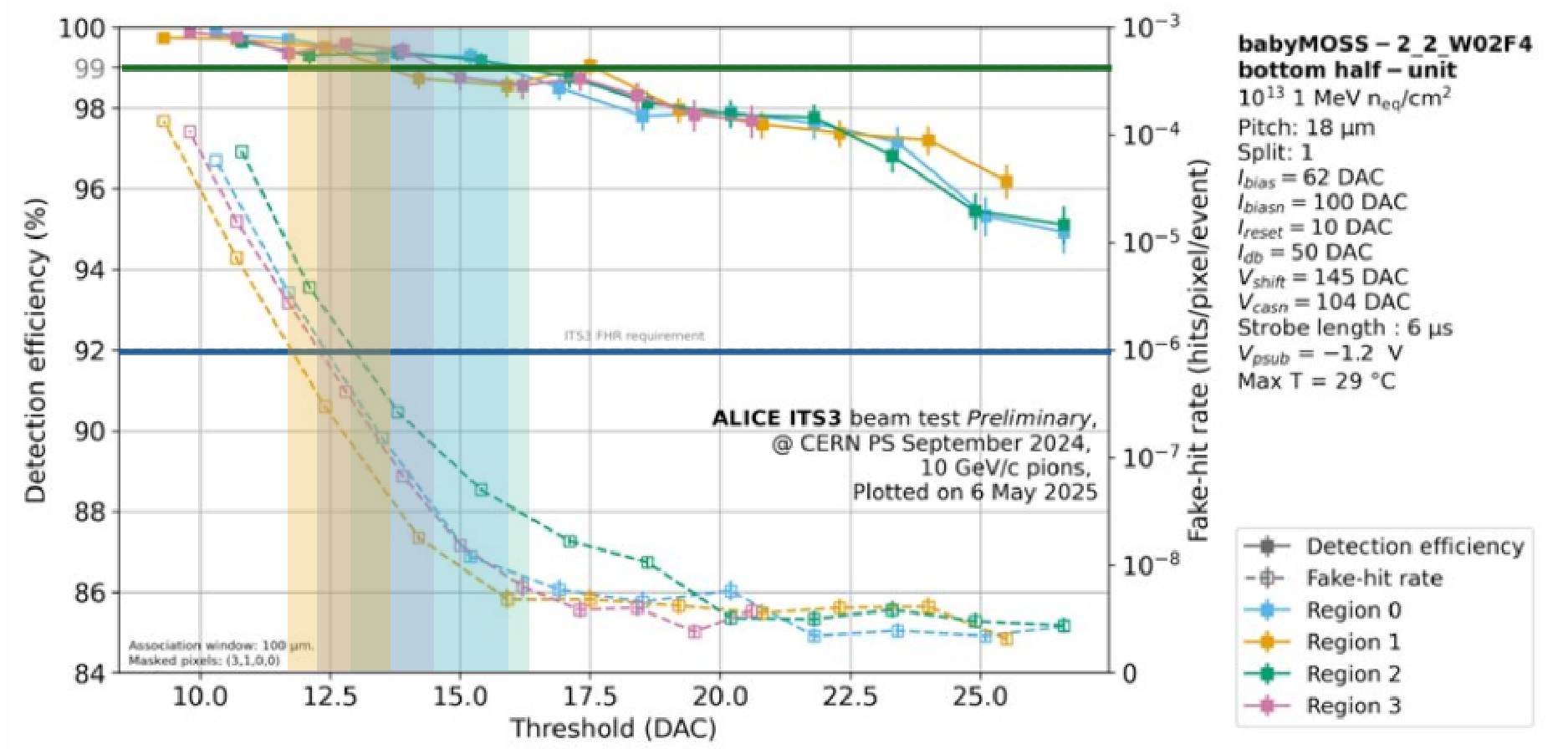}
        \caption{}
        \label{fig:eff_W02F4_bottom}
    \end{subfigure}
    \caption{Detection efficiency and FHR for the irradiated device for Top HU (\subref{fig:eff_W02F4_top}) and Bottom HU (\subref{fig:eff_W02F4_bottom}).}
    \label{fig:eff_W02F4}
\end{figure}

An operational window satisfying the ITS3 requirements with efficiency $> 99\%$ and FHR $< 10^{-6}$ hits/pixel/event is observed for all four regions in both Top and Bottom HUs.
The irradiated device (figure \ref{fig:eff_W02F4}) also satisfies the requirements, and a sufficient operational margin is preserved, confirming the suitability of the sensor for the ITS3 environment.

\subsection{Spatial resolution}
\label{subsec:resolution}
\begin{figure}[ht]
    \centering
    \begin{subfigure}{0.49\textwidth}
        \centering
        \includegraphics[width=\textwidth]{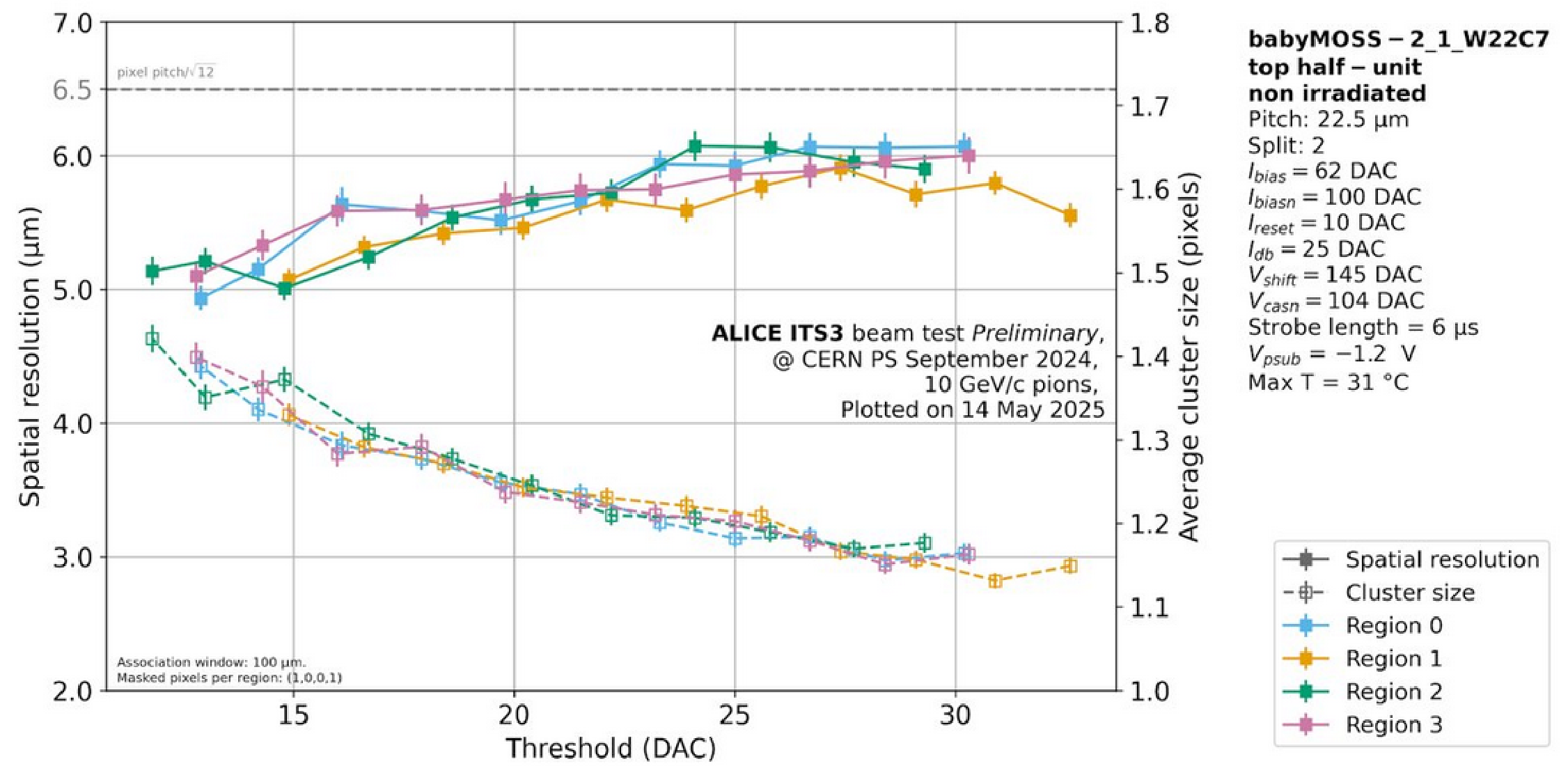}
        \caption{}
        \label{fig:res_W22C7_top}
    \end{subfigure}
    \hfill
    \begin{subfigure}{0.49\textwidth}
        \centering
        \includegraphics[width=\textwidth]{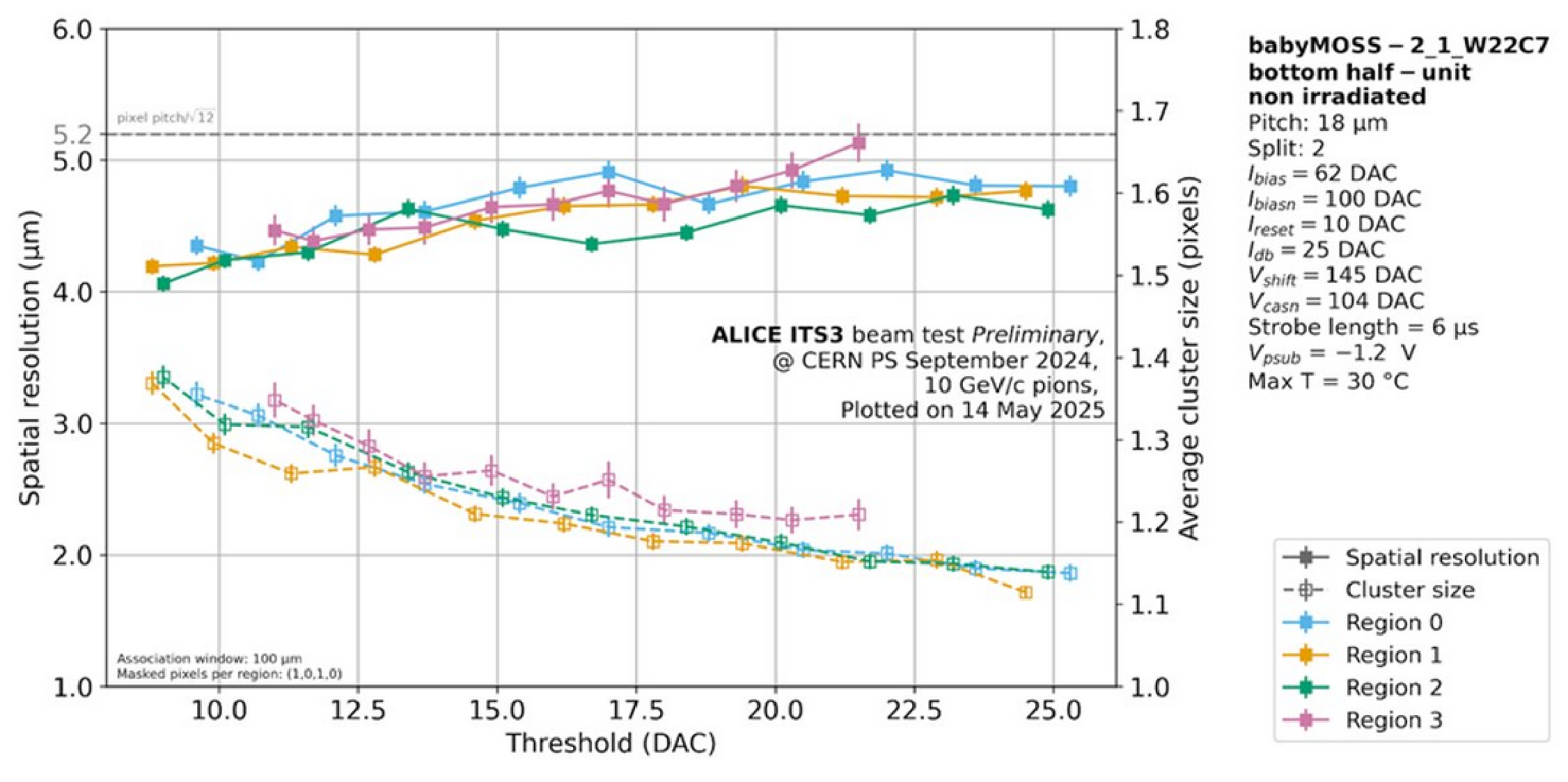}
        \caption{}
        \label{fig:res_W22C7_bottom}
    \end{subfigure}
    \caption{Spatial resolution and average cluster size for the non-irradiated device for Top HU (\subref{fig:res_W22C7_top}) and Bottom HU (\subref{fig:res_W22C7_bottom}).}
    \label{fig:res_W22C7}
\end{figure}

\begin{figure}[ht]
    \centering
    \begin{subfigure}{0.49\textwidth}
        \centering
        \includegraphics[width=\textwidth]{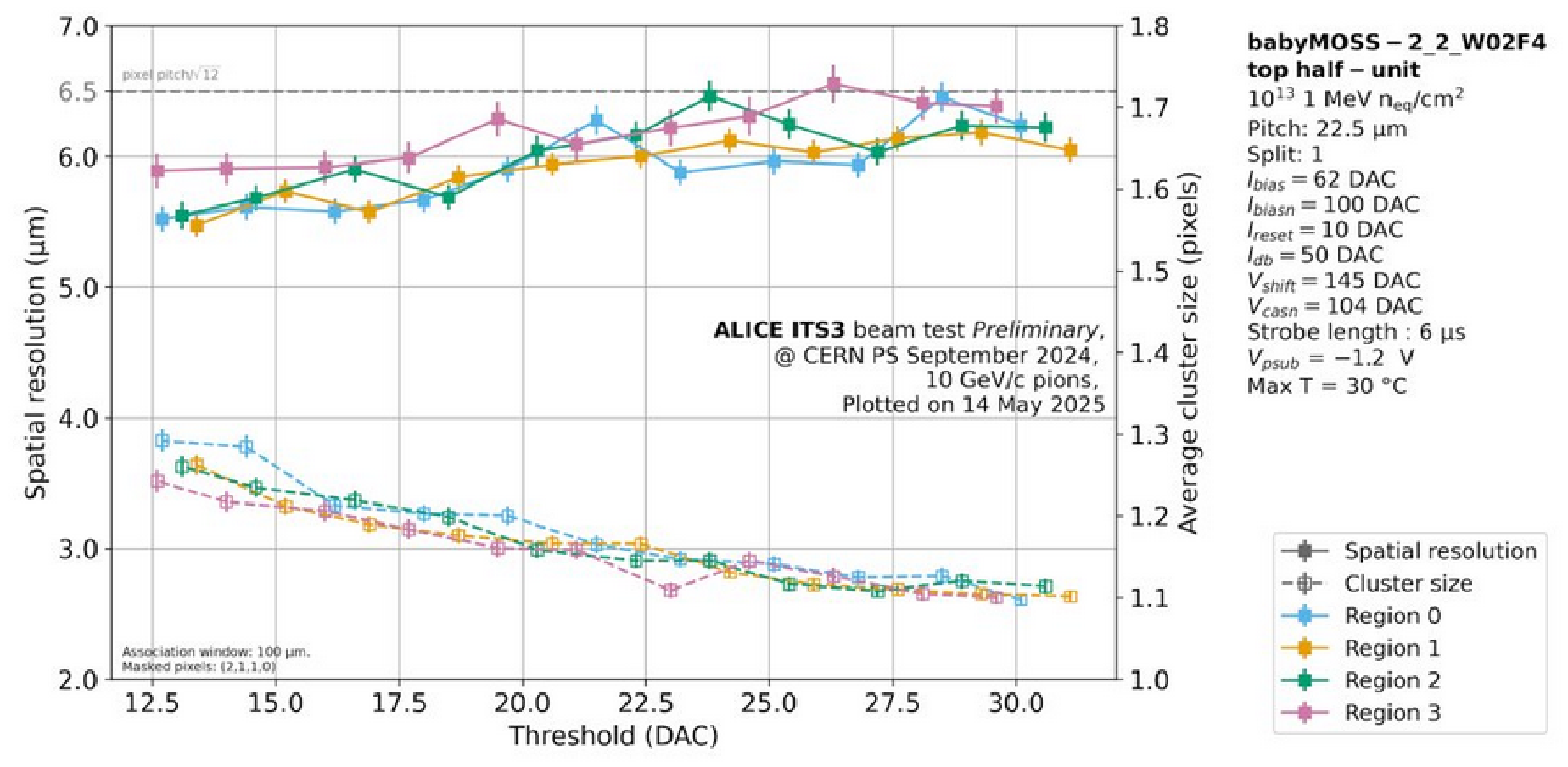}
        \caption{}
        \label{fig:res_W02F4_top}
    \end{subfigure}
    \hfill
    \begin{subfigure}{0.49\textwidth}
        \centering
        \includegraphics[width=\textwidth]{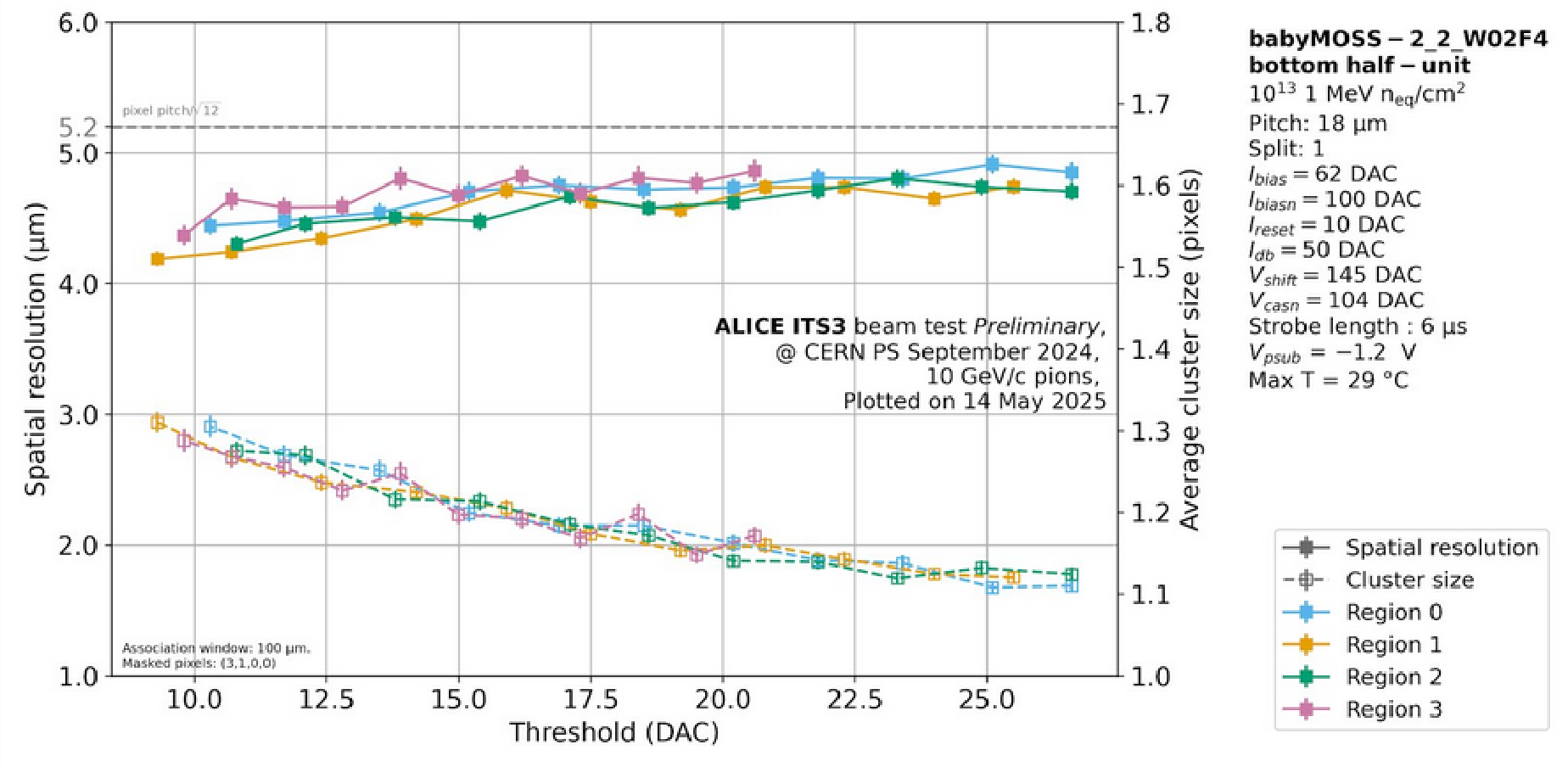}
        \caption{}
        \label{fig:res_W02F4_bottom}
    \end{subfigure}
    \caption{Spatial resolution and average cluster size for the irradiated device for Top HU (\subref{fig:res_W02F4_top}) and Bottom HU (\subref{fig:res_W02F4_bottom}).}
    \label{fig:res_W02F4}
\end{figure}

The Spatial resolution and average cluster size as a function of the threshold are shown in figure \ref{fig:res_W22C7} for the non-irradiated device and in figure \ref{fig:res_W02F4} for the irradiated one. 

The spatial resolution is computed as the average of the width of the residual distributions along the two directions and the resolution along each direction is computed as shown in equation \ref{eq:res}

\begin{eqnletter}
    \label{eq:res}
    \sigma_{i,DUT} = \sqrt{\sigma_{i,fit}^2 - \sigma_{telescope}^2} \quad i=x,y
\end{eqnletter}

where $\sigma_{i,fit}$ is obtained by fitting the residual distributions with a Gaussian function, and $\sigma_{telescope}$ accounts for the tracking precision of the telescope, and it is computed using the Telescope Optimizer tool \cite{telescope_optimizer}.

At low threshold charge sharing increases, leading to a higher average cluster size and therefore a better spatial resolution. Both the non-irradiated and the irradiated chips achieved a spatial resolution better than the binary resolution, equal to \SI{6.5}{\micro\meter} for the Top HU and to \SI{5.2}{\micro\meter} for the Bottom HU. As it can be observed from figure \ref{fig:res_W02F4_top} and figure \ref{fig:res_W02F4_bottom}, the irradiated device exhibits a smaller average cluster size and a slightly worst spatial resolution compared to the non-irradiated device.

\section{Conclusions}
\label{sec:conclusion}
The characterization of the babyMOSS prototypes from ER1 has successfully demonstrated the viability of the stitched 65 nm CMOS sensor technology for the ALICE ITS3 upgrade. The test beam campaign performed at the CERN PS confirmed that the sensors function reliably even at the ITS3 radiation levels specified in section \ref{sec:introduction}.

The two babyMOSS chips characterized exhibited a good operational margin with detection efficiency $> 99\%$ and a fake hit rate $< 10^{-6}$ hits/pixel/event. The spatial resolution is around \SI{5}{\micro\meter}, meeting the requirements set for ITS3.

\end{document}